\documentclass[reprint,letterpaper,amssymb,superscriptaddress,nobibnotes,aps,pra]{revtex4-1}
\usepackage{graphicx}
\usepackage{bm}
\begin{document}

\title{Trapping $^{171}$Yb Atoms into a One-Dimensional Optical Lattice with a Small Waist}

\author{Akio Kawasaki}
\email{akiok@stanford.edu}
\thanks{These authors contributed equally to this work.}
\altaffiliation{Current address: W. W. Hansen Experimental Physics Laboratory and Department of Physics, Stanford University, Stanford, California 94305, USA}
\affiliation{Department of Physics, MIT-Harvard Center for Ultracold Atoms and Research Laboratory of Electronics, Massachusetts Institute of Technology, Cambridge, Massachusetts 02139, USA}

\author{Boris Braverman}
\email{bbraverm@uottawa.ca}
\thanks{These authors contributed equally to this work.}
\altaffiliation{Current address: Department of Physics and Max Planck Centre for Extreme and Quantum Photonics, University of Ottawa, 25 Templeton Street, Ottawa, Ontario K1N 6N5, Canada}
\affiliation{Department of Physics, MIT-Harvard Center for Ultracold Atoms and Research Laboratory of Electronics, Massachusetts Institute of Technology, Cambridge, Massachusetts 02139, USA}

\author{Edwin Pedrozo-Pe\~nafiel}%
\affiliation{Department of Physics, MIT-Harvard Center for Ultracold Atoms and Research Laboratory of Electronics, Massachusetts Institute of Technology, Cambridge, Massachusetts 02139, USA}

\author{Chi Shu}%
\affiliation{Department of Physics, MIT-Harvard Center for Ultracold Atoms and Research Laboratory of Electronics, Massachusetts Institute of Technology, Cambridge, Massachusetts 02139, USA}
\affiliation{Department of Physics, Harvard University, Cambridge, Massachusetts 02138, USA}

\author{Simone Colombo}%
\affiliation{Department of Physics, MIT-Harvard Center for Ultracold Atoms and Research Laboratory of Electronics, Massachusetts Institute of Technology, Cambridge, Massachusetts 02139, USA}

\author{Zeyang Li}%
\affiliation{Department of Physics, MIT-Harvard Center for Ultracold Atoms and Research Laboratory of Electronics, Massachusetts Institute of Technology, Cambridge, Massachusetts 02139, USA}

\author{Vladan Vuleti${\rm{\acute{c}}}$ }
\affiliation{Department of Physics, MIT-Harvard Center for Ultracold Atoms and Research Laboratory of Electronics, Massachusetts Institute of Technology, Cambridge, Massachusetts 02139, USA}

\begin{abstract}
In most experiments with atoms trapped in optical lattices, the transverse size of the optical lattice beams is of the order of tens of micrometers, and loading many atoms into smaller optical lattices has not been carefully investigated. We report trapping 1500 $^{171}$Yb atoms in a one-dimensional optical lattice generated by a narrow cavity mode at a distance of 0.14 mm from a mirror surface. The simplest approach of loading atoms from a mirror magneto-optical trap overlapped with the cavity mode allows the adjustment of the loading position by tuning a uniform bias magnetic field. The number of atoms trapped in the optical lattice exhibits two local maxima for different lattice depths, with a global maximum in the deeper lattice. These results open a way to quantum mechanical manipulation of atoms based on strong interaction with a tightly focused light field. 
\end{abstract}

\maketitle

\section{Introduction}
Optical trapping is widely used in various fields ranging from physics to biology \cite{Nature.424.810} as a way to spatially confine dielectric objects. Optical tweezers were the first realization of this idea \cite{ApplPhysLett.19.283}, where $\sim10$-$\mu$m-sized particles were levitated in low vacuum by the radiation pressure of a single vertically-propagating laser beam. For dielectric particles smaller than a trapping light wavelength, the origin of the potential is the electric polarization induced by the light field, which makes the particle's energy lower at positions with more intense light, when the trapping light is red-detuned from electromagnetic resonant frequency. In the context of atomic physics, this effect is equivalent to the ac Stark shift. Optical trapping can be applied in various ways, including dipole traps originating from a single beam \cite{PhysRevLett.57.314}, optical lattices generated by two \cite{PhysRevLett.68.3861,PhysRevLett.69.49} or more \cite{PhysRevA.50.5173,NewJPhys.12.065025} interfering laser beams, and more complicated arrangements of trap centers by holographically interfering beams \cite{NatCommun.7.13317,PhysRevX.4.021034}. These traps are employed in a variety of experiments, such as optical tweezers (dipole traps) for single atoms \cite{PhysRevX.2.041014} and atom arrays \cite{Science.354.1021,Science.354.1024} and optical lattices for Bose-Einstein condensates \cite{Science.282.1686,PhysRevLett.87.160405}, degenerate Fermi gases \cite{PhysRevA.68.011601}, optical lattice clocks \cite{PhysRevLett.91.223001}, and cavity quantum electrodynamics (cQED) \cite{PhysRevA.69.051804}. Atoms are loaded into these optical traps directly from a magneto-optical trap (MOT) \cite{PhysRevLett.59.2631} or another trap such as a magnetic trap \cite{PhysRevLett.54.2596}. 

The efficiency of loading atoms from a MOT into an optical lattice depends on the relative size of the MOT and the optical lattice. Typically, a MOT has a radius of at least tens of micrometers \cite{PhysRevLett.59.2631,PhysRevA.68.011403,PhysRevA.52.1423,JPhysB.48.155302}, and the counter-propagating beams to generate an optical lattice have a diameter of 40 $\mu$m or more \cite{NatPhoton.9.185,1807.11282,PhysRevLett.116.093602,OptLett.13.4005,NatPhys.11.738,Nature.467.68}. In such situations, the loading efficiency is fairly high, thanks to a large spatial overlap between the MOT and the lattice beams. Loading atoms into a tightly focused optical lattice could be expected to have limited efficiency, given the low ratio of the size of the optical lattice to that of the MOT. However, trapping a large number of atoms into a tight optical lattice is desirable in cQED experiments, where a tight trap overlapping with the narrow cavity-mode waist provides strong atom-light interactions necessary for generating exotic atomic states \cite{RevModPhys.90.035005}.

Several methods have been demonstrated to trap a large number of atoms within a tight cavity mode. One method is to first load atoms into an auxiliary trap adjacent to the cavity mode, such as a magnetic trap or another optical lattice, and to move the atoms to overlap spatially with the cavity mode \cite{PhysRevLett.99.213601,PhysRevLett.98.233601}. This guarantees a good initial overlap between the MOT and the auxiliary trap, leading to the transfer of a large number of atoms, as the atom cloud can be compressed after loading into the second trap. However, this approach adds technical complexity in the system for moving atoms into the cavity mode. Furthermore, not all atomic species are magnetically trappable, precluding the use of an auxiliary magnetic trap. When an optical cavity is in a near-confocal configuration, a MOT can be generated in the center of the cavity, and direct loading to a lattice formed by the narrow cavity mode is possible. In this configuration, loading of $\sim10^5$ atoms into the cavity mode with a 16-$\mu$m waist has been reported \cite{1809.02114}. However, in that experiment, the broad distribution of the atoms along the axial direction of the cavity over as much as 500 $\mu$m lowers the average single-atom cooperativity down to 10\% of the maximum single-atom cooperativity expected for atoms at the cavity waist, reducing the utility of this approach if the goal is maximizing the atom-light interaction. 

In the case of an asymmetric micromirror cavity \cite{PhysRevA.99.013437}, a MOT can also be created directly between the two cavity mirrors. To put atoms in the region of the narrowest part of the cavity mode and thus the strongest atom-light interaction, atoms in the MOT have to be very close to the mirror surface, which potentially shortens the lifetime of atoms in the MOT. This hinders efficient loading and risks contaminating the mirror surface with atoms, decreasing the finesse of the cavity and thus the strength of atom-light interaction \cite{1802.08499}. Loading of atoms into a MOT near a mirror surface itself has been previously demonstrated using a mirror MOT \cite{PhysRevLett.83.3398}, but overlapping the MOT with a small cavity mode further increases the complexity of the experiment. Also, the smallest distance between the atoms and the mirror reported in Ref. \cite{PhysRevLett.83.3398} is not small enough for atoms to reach the narrowest part of the cavity mode described in Ref. \cite{PhysRevA.99.013437}.

In this paper, we demonstrate the loading of atoms into a narrow-waist optical lattice near the surface of a mirror comprising one of the two mirrors in a high-finesse cavity.  The loading efficiency and the properties of the trap are discussed. It is also shown that the increase in the optical loss of the cavity is slow enough to maintain a finesse above $10^4$ for wavelength $\lambda=556$ nm for several years. The advantage of this setup is that we directly load a mirror MOT into the cavity volume, and easily manipulate the atoms' loading position into the intra-cavity optical lattice using a bias magnetic field to move the zero location of the quadrupole magnetic field. 

We use $^{171}$Yb atoms in our cQED apparatus. Ytterbium-171 is one of the leading candidates for the next generation of optical lattice clocks \cite{1807.11282,RevModPhys.87.637}. To further enhance these clocks' precision, it is desirable to utilize entangled states \cite{PhysRevLett.122.223203}, which requires the loading of atoms into the mode volume of optical cavities with a tight waist. Furthermore, Rydberg states \cite{PhysRevA.89.023411,JPhysB.44.184010,1912.08754} can be used to tune the strength of the atom-atom interaction. The results we obtain are applicable to atomic species that have narrow transitions suitable for generating a compact MOT, including alkaline earth and alkaline earth-like atoms, dysprosium \cite{PhysRevLett.107.190401,PhysRevA.82.043425}, erbium \cite{PhysRevLett.96.143005,PhysRevA.85.051401}, thulium \cite{QuantumElectron.44.515}, and alkali atoms when cooled on a narrow transition \cite{PhysRevA.84.063420}.

\section{Mirror MOT for Ytterbium}\label{YbMOT}
\begin{figure}[!tb]
	\begin{center}
\includegraphics[width=0.9\columnwidth]{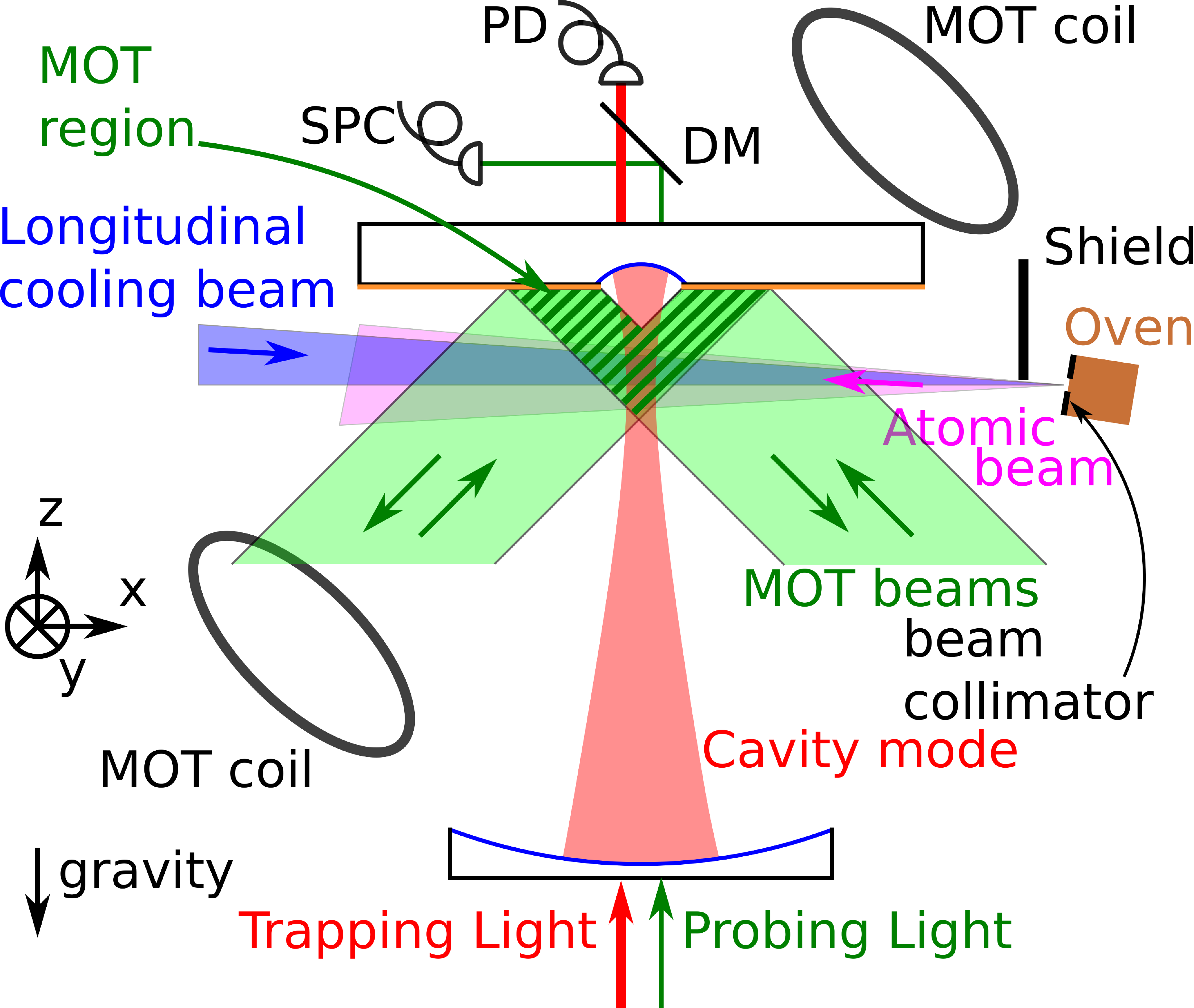}
 \caption{Experimental setup: the diagonal MOT beam is shown in green and the overlapping region of the two diagonal bands is the MOT region where the horizontal beam in the $y$ direction (not shown) also overlaps. The atomic beam from the oven goes through a collimating hole and a shield to protect the mirror surface from being coated directly and is directed towards a point $\sim 5$ mm below the center of the flat mirror with an angle of incidence to the mirror of $\sim80^{\circ}$.  The flat mirror for generating the mirror MOT has a small elliptical pit in the middle, which forms an asymmetric cavity together with a spherical mirror located at $\sim25$ mm below the flat mirror \cite{PhysRevA.99.013437}. SPC, single-photon counter; PD, photodiode; DM, dichroic mirror.} 
 \label{ExpSetup}
 \end{center}
\end{figure}

The experimental setup is shown in Fig. \ref{ExpSetup}. The MOT is in a mirror MOT configuration \cite{PhysRevLett.83.3398}, where a rectangular flat substrate of dimensions 12 $\times$ 25 mm serves as the mirror. Laser beams for the MOT are sent from two directions: horizontally and diagonally. The horizontal beam propagates in the $y$ direction, and the diagonal beam has an angle of incidence to the mirror of $46^{\circ}$ in the $x-z$ plane,
both being retroreflected at the output side. The MOT beams initially have a 1-cm-radius circular shape, and appropriately shaped apertures minimize random scattering on the structure surrounding the mirrors by blocking parts of the MOT beams. MOT coils are on the same axis as the incoming diagonal beam. The atomic beam is produced by an oven $\sim5$ cm away from the MOT region, providing a total atom flux of $\sim10^{10}$ s$^{-1}$ at an oven temperature of $\sim700$ K. 

From the MOT, we load the atoms into a 759-nm optical lattice formed within a high-finesse cavity. The cavity is formed by a micromirror of $\sim150$-$\mu$m diameter, and $\sim344$-$\mu$m radius of curvature (ROC) fabricated into the flat substrate \cite{NJP12.065038} on one side and a 25-mm ROC mirror located 25.0467(10) mm below the flat mirror substrate on the other side (see Ref. \cite{PhysRevA.99.013437} for additional details on the cavity structure). For 759-nm light used for optical trapping, the $1/e^2$ beam radius at the waist of the cavity mode is $w_0=5.4$ $\mu$m, and the Rayleigh range is 0.12 mm. This gives a beam waist of 6.6 $\mu$m at $Z=0.14$ mm, where $Z$ is the distance of the atoms from the micromirror substrate. 

Because the mirror MOT geometry and the structure of the vacuum chamber limit the maximum quadrupole magnetic-field gradient, we use a two-color MOT (TCMOT) \cite{JPhysB.48.155302} to trap $^{171}$Yb, where 399- and 556-nm lasers near resonant to the $^1S_0 \rightarrow ^1$$P_1$ transition and the $^1S_0 \rightarrow ^3$$P_1$ transition, respectively, are used simultaneously (see Ref. \cite{JPhysB.48.155302} for the energy level diagram). Properties of the transitions and parameters of the MOT beams for the TCMOT are summarized in Table \ref{TCMOTparameters}. In addition to the MOT beams, we apply a longitudinal cooling beam counterpropagating to the atomic beam with a power of 1.8 mW and a detuning of $-4.6\Gamma_{\rm s}$ from the $^1S_0 \rightarrow ^1$$P_1$ transition, which has a radius of 1 mm in the MOT region. With a magnetic-field gradient of 14.4 G/cm, typically $10^4$ atoms are trapped in the TCMOT. 

\begin{table}[!t]
\caption{Properties of the two transitions used for the MOT and parameters of the MOT beams for the TCMOT. }
\begin{center}
\begin{tabular}{ccc}
Transition & $^1S_0 \rightarrow ^1$$P_1$ & $^1S_0 \rightarrow ^3$$P_1$ \\
\hline
Wavelength (nm)  & 399.911 & 556.799  \\
Linewidth$/2\pi$ (MHz) & $\Gamma_{\rm s}=29.1$  & $\Gamma_{\rm t}=0.184 $  \\
Saturation intensity (mW/cm$^2$) & $I_{\rm sat, s}=57$  & $I_{\rm sat,t}=0.14$  \\
Laser intensity & $0.10I_{\rm sat,s}$ & $50I_{\rm sat,t}$ \\
Laser detuning & $-0.71\Gamma_{\rm s}$ & $-38\Gamma_{\rm t}$ \\

\end{tabular}
\end{center}
\label{TCMOTparameters}
\end{table}%

\begin{figure}[!tb]
	\begin{center}
 \includegraphics[width=1\columnwidth]{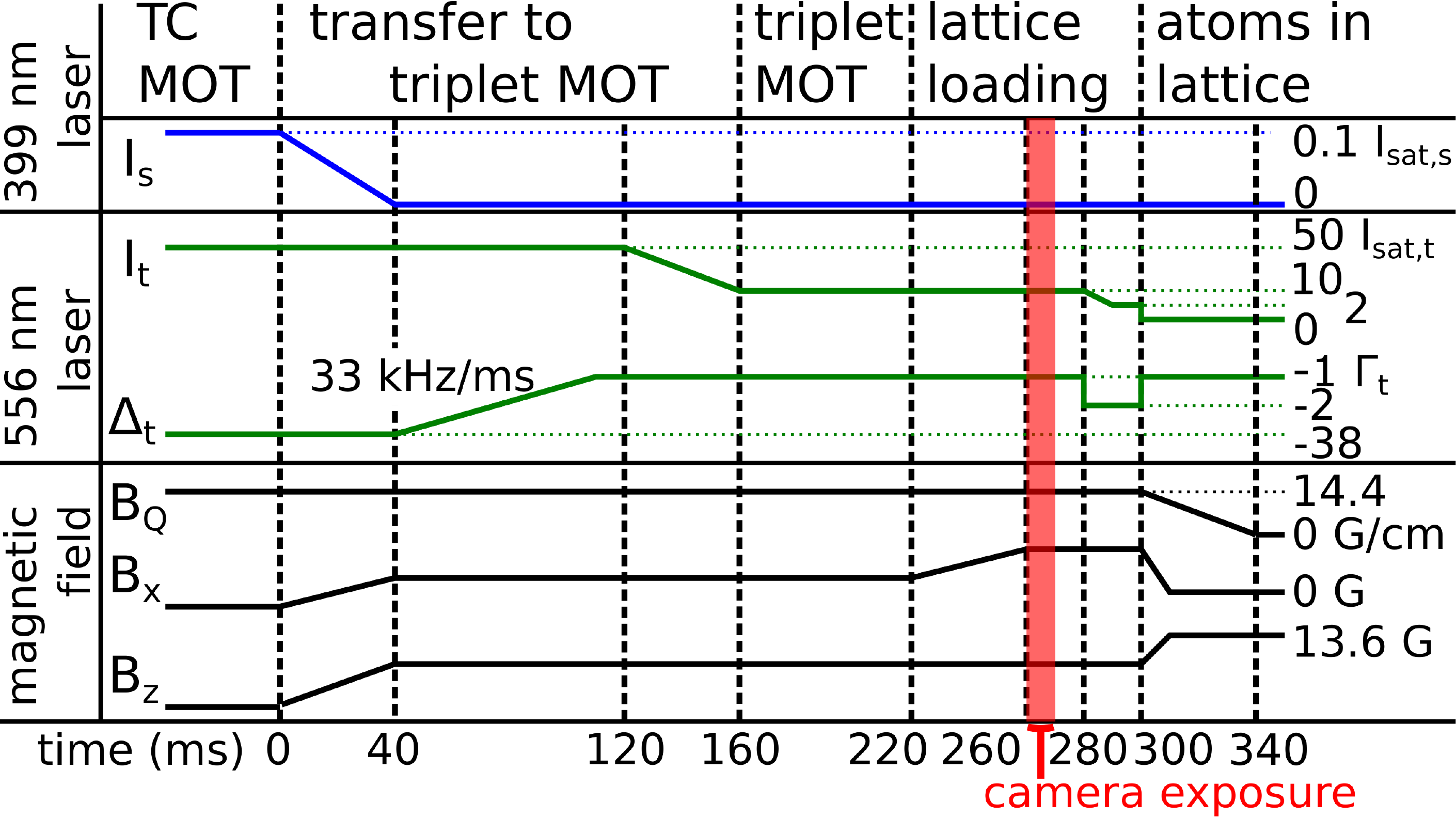}
 \caption{Typical sequence for loading atoms into the optical lattice. Numbers for the bias field $B_{\rm y}$ are not shown because they change according to the loading point and slow drift of the background field but are typically a few gauss. $I_{\rm s}$, $I_{\rm t}$, $\Delta_{t}$, $B_{\rm Q}$, $B_{\rm x}$, and $B_{\rm z}$ are the intensity of the 399-nm laser, intensity and detuning of the 556-nm laser, quadrupole magnetic-field gradient along the strong-field axis, bias magnetic field in the $x$ direction, and bias magnetic field in the $z$ direction, respectively.} 
 \label{LoadingSequence}
 \end{center}
\end{figure}

Once the TCMOT loading stage of a few seconds is completed, atoms are transferred from the TCMOT into a MOT with only the 556-nm light (triplet MOT) \cite{PhysRevA.60.R745}. The sequence to generate the triplet MOT, which is optimized to maximize the transfer efficiency from the TCMOT to the triplet MOT, is summarized in Fig. \ref{LoadingSequence}. The gradual changes in parameters maintain a larger number of atoms in the MOT, compared to sudden step-function-like quenches. Note that tuning the $x$ direction bias magnetic field actually moves the atom position along the $z$ axis, due to the tilted quadrupole field from the MOT coils.

The temperature of atoms in the triplet MOT is measured by the time-of-flight method with absorption imaging by a CCD camera using a 399-nm laser resonant with the $^1S_0 \rightarrow ^1$$P_1$ transition, sent horizontally at a $\sim15^{\circ}$ angle relative to the $y$ axis. The temperature of the atoms in the triplet MOT is $\sim15$ $\mu$K. 

\begin{figure}[!tb]

\includegraphics[width=0.7\columnwidth]{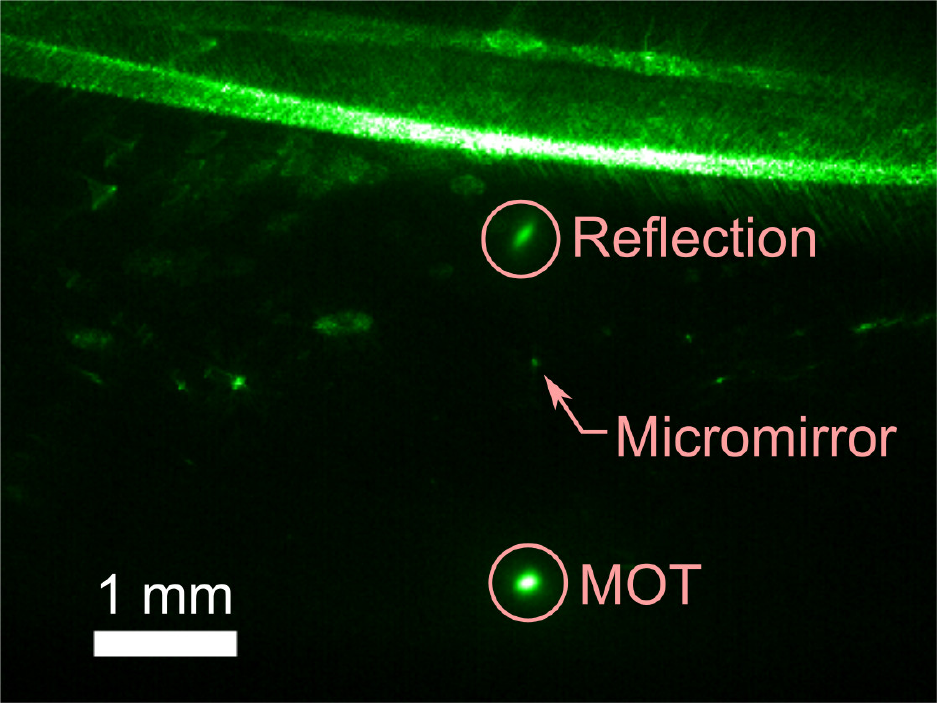}
 \caption{Image of a triplet MOT and its reflection from the flat mirror surface. 
Light scattered from imperfections in the flat mirror substrate is also visible in this image. The arrow marks the position of the micromirror, which also scatters some light. The MOT is 1.28 mm away from the mirror surface. The edge of the micromirror substrate is outside the field of view of the camera. } 
 \label{MOTPhoto}

\end{figure}

The atom position $Z$ relative to the flat mirror is measured by a CCD camera imaging along an angled direction, with a tilt of $14.2\pm0.2^{\circ}$ to the plane of the mirror. Both a direct image and a reflected image in the mirror of the MOT are visible in the image acquired by the camera when the MOT is sufficiently close to the mirror (Fig. \ref{MOTPhoto}). $Z$ is calculated from the separation $d$ between the two images of the MOT as $Z=d/(2\cos (14.2^{\circ}))$. To remove image artifacts due to significant amount of light scattered from the mirror substrate by surface defects, background is subtracted by acquiring a reference image at the end of each experimental sequence, after removing all remaining atoms with a pulse of 399-nm light. $Z$ is affected by the detuning of the 556-nm laser from the atomic resonance due to the influence of gravity and imbalance between the incident and reflected MOT beam intensities. With a fixed 556-nm laser frequency, the triplet MOT position is stable within 10 $\mu$m in experimental runs spanning an hour. We ascribe the position drift to slow ambient magnetic-field fluctuations of the order of $\pm 10$~mG, estimated from changes in the ground-state Larmor frequency.

\begin{figure}[!tb]
	\begin{center}
\includegraphics[width=1\columnwidth,bb=0 0 2400 1800]{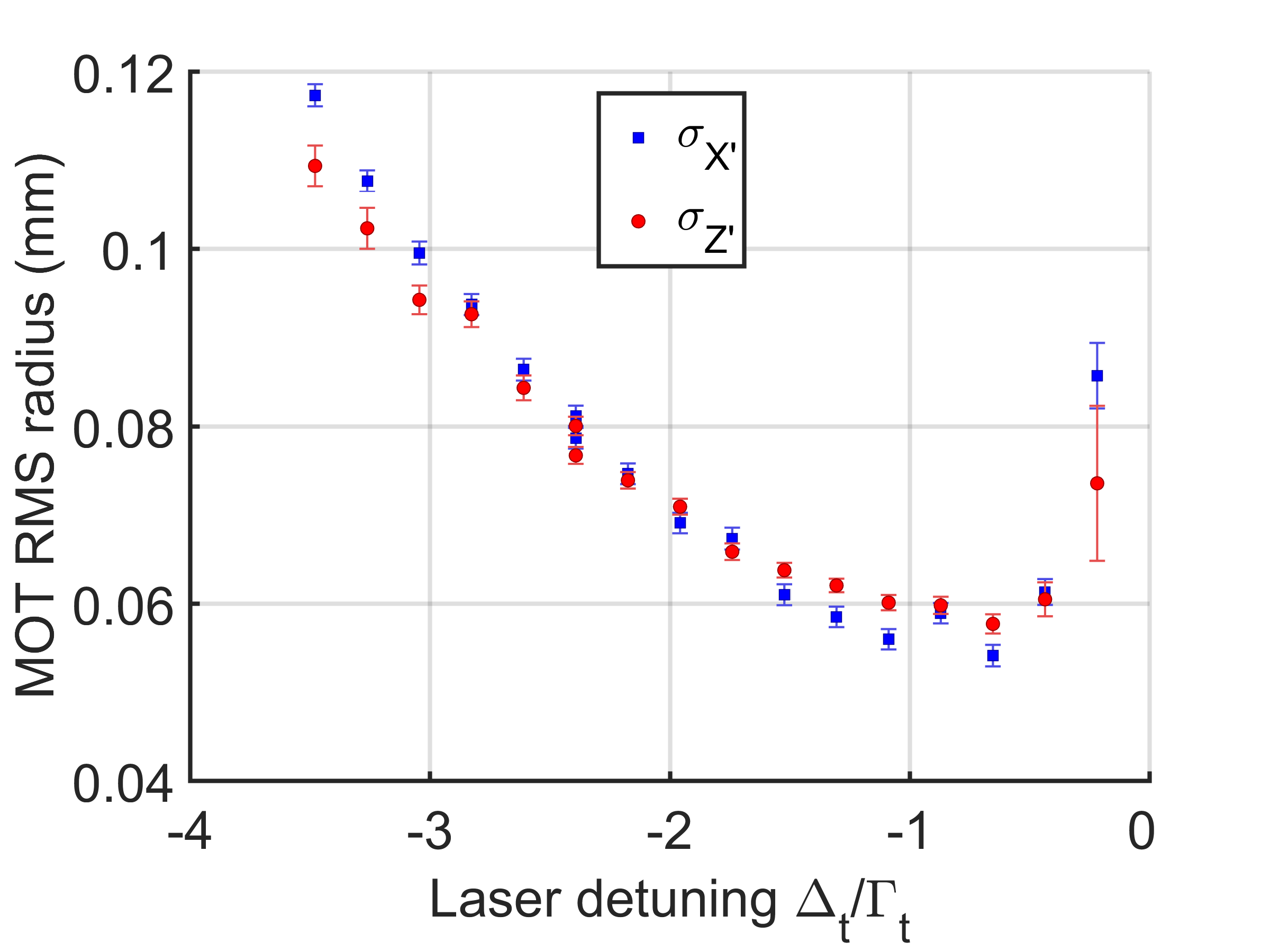}
 \caption{The size of the triplet MOT for different detunings of the 556-nm laser from resonance: at the typical detuning of $\Delta=-\Gamma_{\rm t}$, the MOT is well compressed to 55-$\mu$m RMS radius.} 
 \label{Fig5-2}
\end{center}
\end{figure}

The imaging of the triplet MOT also gives the size of the MOT, as shown in Fig. \ref{Fig5-2}. A detuning of $\Delta=-\Gamma_{\rm t}$ in the cooling stage compresses the MOT down to a root-mean-square (RMS) radius of 55 $\mu$m and optimizes the lattice loading efficiency.

\begin{figure}[!tb]
	\begin{center}
 \includegraphics[width=1\columnwidth,bb=0 0 1198 775]{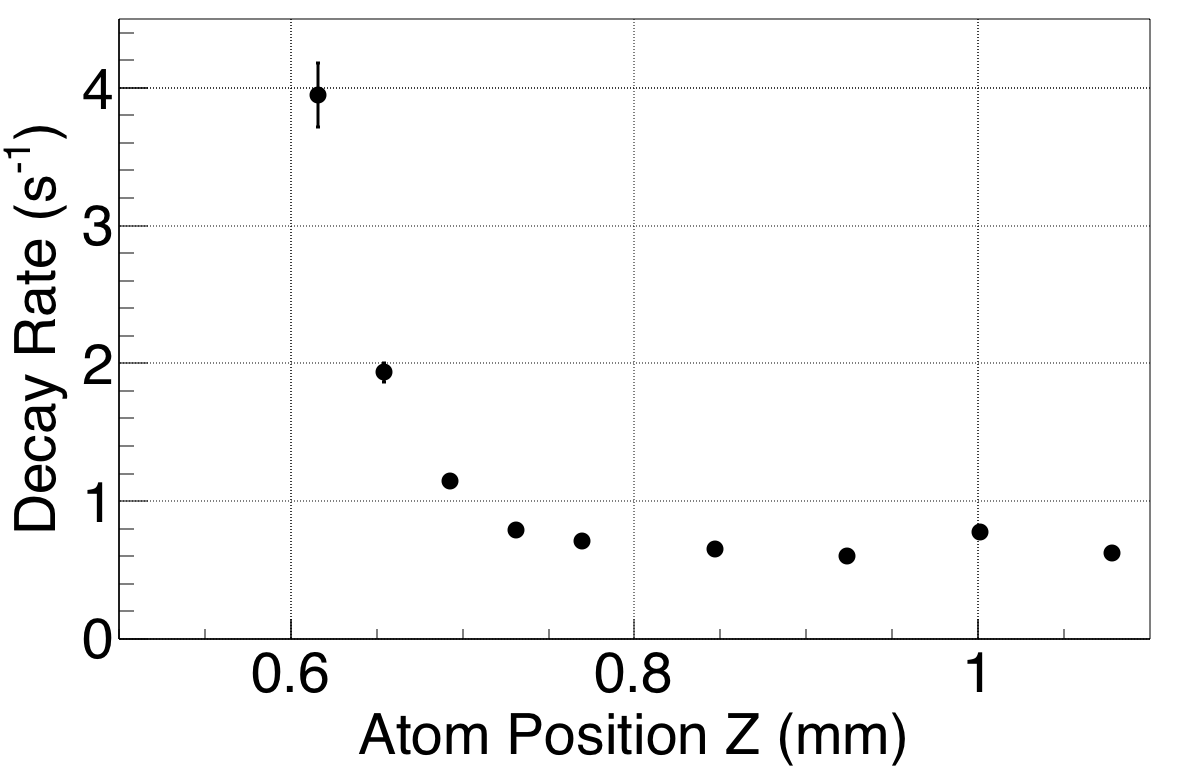}
 \caption{Decay rate of the atom number in the triplet MOT at different distances from the mirror.} 
 \label{DecayRateVSDistance}
 \end{center}
\end{figure}

The lifetime of atoms in the triplet MOT at different positions $Z$ is measured by continuously monitoring the amount of fluorescence from the triplet MOT by an avalanche photodiode, and extracting the exponential decay rate from the total fluorescence of the MOT. Figure \ref{DecayRateVSDistance} shows the change in the triplet MOT lifetime according to the distance from the mirror. The decay rate increases at $Z\lesssim 0.7$ mm, and the decay becomes too fast to observe at $Z=0.6$ mm. We speculate that the region where atoms can be trapped spreads over 0.7 mm vertically upward from the center of the MOT. This is reasonably consistent with Ref. \cite{ApplPhysB.74.469}, which shows a sharp increase in the loss rate below a certain MOT-surface distance. The mechanism of the larger loss when the MOT is closer to the mirror is presumably that the outer regions of the MOT overlap with the mirror and atoms stick to the mirror. Note that this measurement is performed at the magnetic-field gradient of 9 G/cm. With a larger magnetic-field gradient of 14.4 G/cm and repeated optimization over time of the beam alignment and polarizations, atoms are trapped by the triplet MOT for a few hundreds of milliseconds at $Z=0.14$ mm.

\section{Loading to the optical lattice}\label{LatticeLoading}
The loading sequence into an optical lattice made of 759-nm light is summarized in Fig. \ref{LoadingSequence}. As the 759-nm trap light is always circulating in the cavity mode (see Sect. \ref{opticalpumping}), simply turning off the 556-nm laser with atoms at the desired location transfers atoms to the optical lattice. In the last 20 ms of the triplet MOT, the intensity of the 556-nm laser is reduced to $2I_{\rm sat,t}$, in addition to an increase in the detuning to $-2.7\Gamma_{\rm t}$. 
This larger red-detuning is necessary to compensate for the ac Stark shift induced by the trapping light, which is 25\% larger for the $^3$P$_1$ excited state than the $^1$S$_0$ ground state. 
(It is assumed that information on the ac Stark shift for $^{174}$Yb in Ref. \cite{JPhysB.51.125002} is the same for $^{171}$Yb when averaged over the hyperfine structure.) The quadrupole magnetic field is kept constant over the entire sequence up to this point. 

After the 556-nm laser is turned off instantaneously, the quadrupole magnetic field and radial bias fields $B_x, B_y$ are ramped down to 0, while the axial magnetic field $B_z$ is ramped to a specific value, typically 13.6 G. The turning-off of the quadrupole magnetic field is performed gradually over 40 ms, to avoid a mechanical kick to the cavity structure leading to oscillations that prevent reliable probing of the cavity resonance.

\section{Optical lattice properties}\label{opticalpumping}
The one-dimensional optical lattice consists of 759-nm light circulating in the cavity mode, with a finesse of $3.14(6)\times 10^3$ at 759 nm. Because of imperfect mode matching between the input light and the cavity mode and losses at the mirror surfaces, the coupling efficiency of the input light to the cavity mode is limited to 19\%. An input power of $P_{759}=9.9$ mW of the 759-nm laser to the cavity therefore generates an optical lattice in the cavity equivalent to a 1.92-W retroreflected beam. The cavity is locked to the 759-nm laser by Pound-Drever-Hall (PDH) technique \cite{ApplPhysB.31.97} to ensure that the power enhancement is always present, as well as to maintain the cavity resonant frequency at a specific value. The 759-nm light is generated by a distributed Bragg reflector laser that is PDH locked to a separate stable reference cavity. To reduce the heating of atoms by intensity fluctuations of the optical lattice converted by the cavity from laser frequency fluctuations, the laser has an electro-optic modulator (EOM) feedback system that reduces the linewidth down to $\sim 1$ kHz \cite{EOMFeedbackLaser}. 

The trapping frequency of the optical lattice is measured by modulating the intensity of the trap laser. The typical time scale $T$ for parametric heating of atoms in the lattice obeys the formula \cite{PhysRevA.56.R1095}
\begin{equation}\label{latticelifetimeeq}
\frac{1}{T}=\pi^2 \nu^2 S(2\nu),
\end{equation}
where $\nu$ is the trapping frequency, and $S$ is the relative intensity noise of the trap laser. According to this formula, when we intentionally modulate the intensity of the lattice beam at a frequency twice as high as the trapping frequency, $T$ decreases significantly compared to when the intensity is modulated at a different frequency. 
The population of atoms in the trap is measured at $Z=1.99$ mm by the dispersive shift of the cavity resonant frequency \cite{AKThesis,BBThesis,PhysRevA.99.013437}, where the measured atom number decay is fitted by an exponential function to extract $T$. Measured radial and axial trapping frequencies are $\nu_{\rm r}=125\pm 5$ Hz (radial) and $\nu_{\rm ax}=67\pm 2$ kHz (axial). 
Using the following equations to calculate $\nu_{\rm ax}$ and $\nu_{\rm r}$ in terms of the trap depth $U_0$, the atom mass $m$, the wave number $k$, and the waist size $w$,
\begin{eqnarray}
\nu_{\rm ax}&=&\frac{1}{2\pi}\sqrt{\frac{2U_0k^2}{m}} \label{OmegaAx}\\
\nu_{\rm r}&=&\frac{1}{2\pi}\sqrt{\frac{4U_0}{w^2m}} \label{OmegaRad}, 
\end{eqnarray} 
the trap depth is estimated to be $U_0= h \times 554~{\rm kHz}=277~E_{\rm r}$, where $E_{\rm r}=h \times 2.02$ kHz is the recoil energy, at $Z=1.99$ mm. The value is reasonably consistent with expected values from the 1.92-W intracavity power and the previously reported trap depth for the $^1$S$_0$ ground state with a trap beam intensity of 100 kW/cm$^2$ \cite{PhysRevLett.100.103002,JPhysB.43.074011}. 

With the EOM feedback to the trapping laser \cite{EOMFeedbackLaser}, the frequency noise is so low that the lifetime of the atoms in the lattice is 2 s or more, which is long enough to perform cQED experiments. The temperature of the atoms in the lattice is $\sim30$ $\mu$K, measured by the motional sideband spectroscopy using the $^1S_0 \rightarrow ^3$$P_0$ clock transition. Compared with the temperature in the triplet MOT ($\sim15$ $\mu$K), the atom temperature in the optical lattice is hotter, which is consistent with the previous observations in ytterbium \cite{PhysRevLett.91.040404,PhysRevLett.98.030401,ChinPhysLett.33.070601} and strontium \cite{PhysRevA.70.063413,PhysRevLett.96.033003} systems.

\section{Loading Efficiency for Different Lattice Powers and MOT-Mirror Distances}\label{LoadEff}
\begin{figure*}[!tb]
	\begin{center}
\includegraphics[width=2\columnwidth]{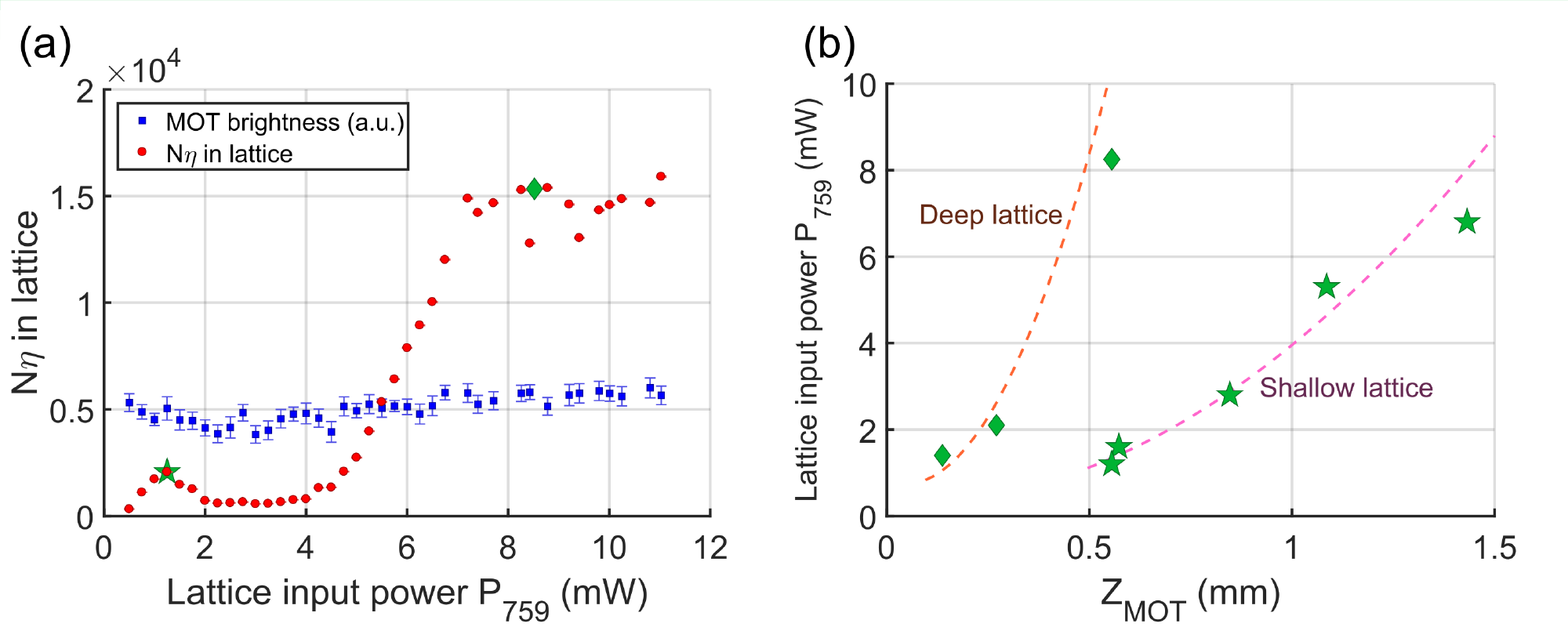}
 \caption{(a) MOT brightness (blue squares) and atom number loaded into the optical lattice (red circles) at a fixed atom position $Z=0.54$ mm, as a function of the input 759-nm laser power. A star around 1 mW and a diamond around 8 mW correspond to the shallow lattice and deep lattice data points in (b), respectively. (b) Values of trap laser power that locally maximize atom loading efficiency. The two points at $Z=0.54$ mm correspond to the maxima shown in (a). Two dashed lines show the laser power needed to provide constant lattice depths for the deep and shallow lattice regimes. The corresponding lattice depths are shown in Fig. \ref{Fig5-5} with the dashed lines of the same colors.}
 \label{Fig5-4}
\end{center}
\end{figure*}

The atom number trapped in the lattice is measured via interaction with a high-finesse cavity for 556 nm resonant to the $^1S_0 \rightarrow ^3$$P_1$ transition. Therefore, the atom number in the lattice is measured through the collective cooperativity $N\eta$, where $N$ is the atom number and $\eta=24{\cal F}/\pi k^2 w_{556}^2$ is the single-atom cooperativity ($\eta$ through this paper refers to the single-atom cooperativity, not Lamb-Dicke parameter), with the finesse ${\cal F}$, the wave number $k$, and the waist size for the cavity mode of 556-nm light $w_{556}$. In our system, $\eta=1.5$ at $Z=0.54$ mm, and $N\sim 10^4$ (see Fig. 5 in Ref. \cite{PhysRevA.99.013437} for more detailed analysis). 
Figure \ref{Fig5-4}(a) shows $N\eta$ in the lattice, measured by the amount of Rabi splitting \cite{AKThesis,BBThesis,PhysRevA.99.013437} for different input powers $P_{\rm 759}$ at $Z=0.54$ mm. There are two local maxima; a shallow-lattice one is located around $P_{759}=1.4$ mW, and a deep-lattice one is at $P_{759}\geq 7$ mW. When maxima in similar measurements for different $Z$ are plotted [Fig. \ref{Fig5-4}(b)], the deep-lattice regime is only observed for $Z \leq 0.54$ mm, and the shallow lattice is seen at $Z \geq 0.54$ mm. We expect that it is also possible to observe deep-lattice regime at larger $Z$, with larger $P_{759}$. Note that the 556 nm laser is detuned by $\Delta_{\rm t}=-500~{\rm kHz}=-2.72 \Gamma_{\rm t}$ from the $^1S_0 \rightarrow ^3$$P_1$ transition during the final cooling stage for all experiments described in this section.

\begin{figure}[!tb]
	\begin{center}
 \includegraphics[width=1\columnwidth]{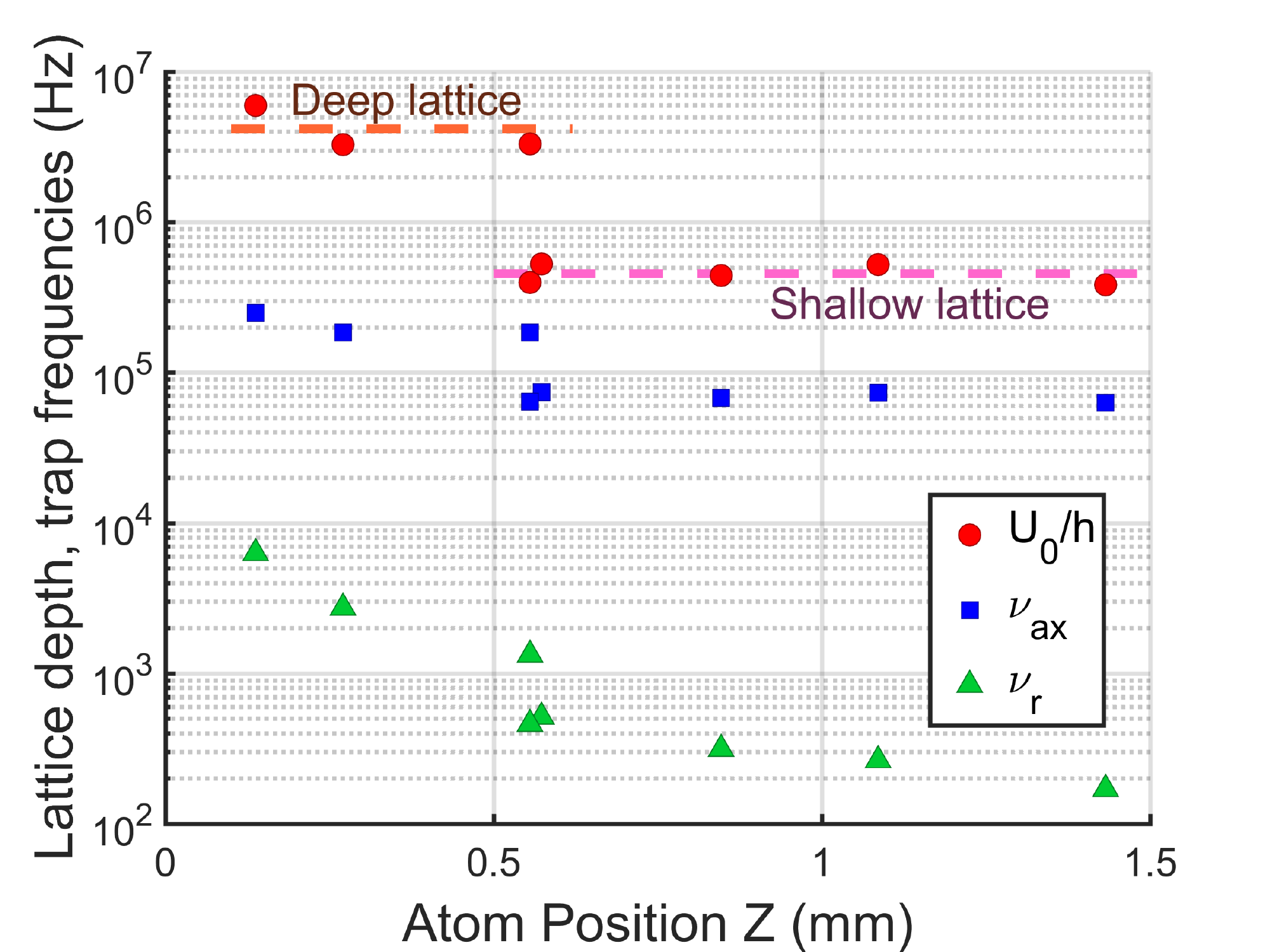}
 \caption{Trap depth $U_0$ (red circles) and axial (blue squares, $\nu_{\rm ax}$) and radial (green triangles, $\nu_{\rm r}$) trapping frequencies of trap depths locally maximizing $N\eta$ in the lattice at different atom position $Z$: $\nu_{\rm r}$ is the geometrical mean of two orthogonal radial directions. The deep and shallow lattice regimes correspond to lattice depths $U_0/h$ of approximately 4.2~MHz and 450~kHz, respectively.} 
 \label{Fig5-5}
\end{center}
\end{figure}

To investigate the mechanisms responsible for two local maxima, we plot the trap depths $U_0$ and trapping frequencies $\nu_{\rm r}$ and $\nu_{\rm ax}$ in Fig. \ref{Fig5-5}, corresponding to the conditions in the measurements in Fig. \ref{Fig5-4}. As expected from Fig. \ref{Fig5-4}, the lattice depth $U_0$ and the axial trapping frequency $\nu_{\rm ax}$ remain constant in both the deep-lattice regime and the shallow-lattice regime. 

In the deep-lattice region, $U_0 \simeq h \times 4$ MHz and $\nu_{\rm ax} \simeq 200 $ kHz $\simeq \Gamma_{\rm t}$, which suggests the presence of sideband cooling \cite{PhysRevLett.81.5768,PhysRevLett.80.4149,EPL.42.395}. Sideband cooling only happens for a sufficiently deep lattice because the trap depth $U_{^{3} \hspace{-0.1 em} P_1}$ for the $^3P_1$ excited state is 25\% larger than the trap depth $U_{^{1} \hspace{-0.1 em} S_0}$ for the $^1S_0$ ground state \cite{JPhysB.51.125002}, primarily due to coupling to the 6s7s$^3S_1$ state. In the deep lattice, the effective detuning of the cooling laser $\Delta_{\rm eff}$ near the bottom of the potential becomes small in magnitude, and sideband cooling becomes possible only when the excitation to a lower motional state is resonant, i.e. $\Delta_{\rm eff}=\Delta_{\rm t} + (U_{^{3} \hspace{-0.1 em} P_1}-U_{^{1} \hspace{-0.1 em} S_0})/h \approx - \nu_{\rm ax}$, where $(U_{^{3} \hspace{-0.1 em} P_1}-U_{^{1} \hspace{-0.1 em} S_0})/h\approx-1$ MHz for $U_{^{1} \hspace{-0.1 em} S_0}/h \simeq 4$ MHz when averaging the shift over the Zeeman sublevels of the $^3P_1$, $m_F = 3/2$ state. This leads to an increased trap loading efficiency as atoms are cooled into the minimum of the trapping potential.

For the shallow-lattice regime, $U_0/h$ is around 400 kHz, while $\nu_{\rm ax}$ is 65 kHz. This roughly means that $U_0/h\approx-\Delta_{\rm t}$, where an atom can dissipate all its excess kinetic energy acquired when moving into a lattice potential minimum by scattering a single photon at the atom's resonance frequency. A detuning $\Delta_{\rm t}$ slightly larger than $U_0$ is optimal for stopping atoms with finite initial kinetic energy. For deeper lattices, this cooling mechanism no longer works and the loading efficiency decreases.

\section{Lattice Loading Near the Micromirror Surface}
A previous study of the mirror MOT \cite{ApplPhysB.74.469} reports that the atom lifetime decreases rapidly as the atoms are brought closer than 0.2 mm to a surface. However, our system has two advantages compared to \cite{ApplPhysB.74.469}:  atoms in the MOT are kept near the mirror surface only for a short time before they are loaded into the lattice, and the triplet MOT is very compact with an RMS radius of 55 $\mu$m, due to the narrow linewidth of the cooling transition. To load the atoms very near the micromirror surface, the MOT location is moved upward by simply ramping the bias magnetic field. Subsequently, the MOT beams are suddenly turned off.

\begin{table}[!t]
\caption{List of previously reported rates of the loss increase in cavity mirrors: d${\mathcal L}$/dt is the rate of the loss increase, $\lambda$ is the wavelength at which the cavity has a high finesse, and RT is room temperature. The number for Ref. \cite{PhysRevA.95.033812} is obtained from a private communication, and can also be read from Fig. 1 of Ref. \cite{PhysRevA.95.033812}. Reference \cite{OptExp.23.18014} concerns a cavity system without atoms, and Refs. \cite{PhysRevA.95.033812,1902.06014} concern cQED systems with Yb$^+$ ions. Micromirrors are used for one of the two mirrors in this work, and both mirrors in Ref. \cite{PhysRevA.95.033812}, while all other mirrors have large radii of curvature. }
\begin{center}
\begin{tabular}{ccccc}
Ref. 							& d${\mathcal L}$/dt (ppb/h) & Temperature & $\lambda$ (nm)	& Top layer \\
\hline
This work						& $6.76\pm0.01$ & 30 $^{\circ}$C	& 556 nm		& SiO$_2$		\\
\cite{PhysRevA.95.033812}		& $2300\pm200$	& RT				& 369 nm		& SiO$_2$		\\
\cite{OptExp.23.18014}			& $12.3\pm4.3$	& 21 $^{\circ}$C	& 370 nm		& Ta$_2$O$_5$	\\
\cite{OptExp.23.18014}			& $230\pm30$	& 100 $^{\circ}$C	& 370 nm		& SiO$_2$		\\
\cite{1902.06014}				& $0.9\pm3.5$	& RT				& 369 nm		& SiO$_2$	

\end{tabular}
\end{center}
\label{LossComp}
\end{table}

As shown in Figs. \ref{Fig5-4} and \ref{Fig5-5}, the smallest distance between the atom loading position in the cavity and the micromirror surface is only 0.14 mm. At this position, $N\eta\approx 10^4$ and $\eta=10$ are observed \cite{PhysRevA.99.013437}, which implies a total atom number $N_{\rm tot}=(3/2)N\simeq1500$ (see Ref. \cite{PhysRevA.92.063816} or Ref. \cite{PhysRevLett.104.073604} for the derivation). The lifetime of atoms in the lattice is shorter when $Z$ is small, estimated to be around 0.5 s for $Z<0.25$ mm, whereas the typical lifetime is more than 1 s for larger $Z$. We believe that the atom lifetime in the trap is shorter for smaller $Z$ primarily due to the stronger probe-induced atom loss during the measurement of the atom number at smaller $Z$, as the single-atom cooperativity $\eta$ is larger for smaller $Z$. Indeed, we observe a linear rather than exponential decay of the atom number with stronger probing, which is consistent with probing-induced loss rather than one-body loss. Here, the atom number decay is linear because the reduction in the loss rate for smaller $N\eta$ is compensated by probe light being closer to atomic resonance due to the reduced vacuum Rabi splitting.

\begin{figure}[!tb]
	\begin{center}
 \includegraphics[width=1\columnwidth,bb=0 0 2400 1800]{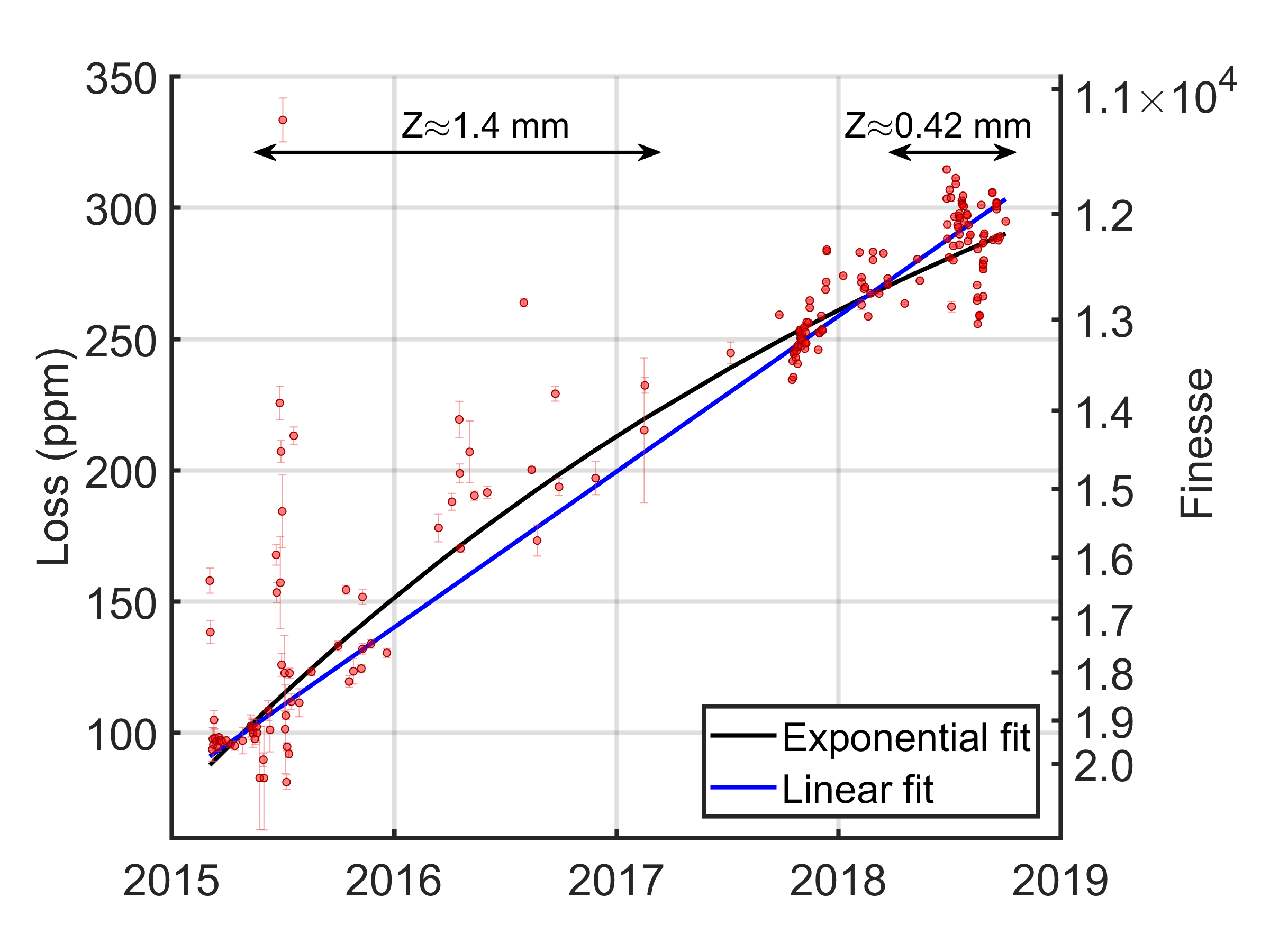}
 \caption{Change in the cavity loss at 556 nm over time: the red dots are the measured values, and the blue line is a linear fit. The fluctuation of the loss away from the trendline significantly exceeding the error bar results from the finesse drift induced by slight changes in the cavity alignment. To avoid large effects from these outliers, Lorentzian weight is used for the fit, instead of the Gaussian weight in the standard $\chi^2$ fit. The slope obtained by the fit is 6.76(1) ppb/h. The black line is a fit by the exponential function ${\mathcal L}={\mathcal L}_0+A(1-\exp(-t/\tau))$ described in \cite{OptExp.23.18014}. The fitted parameters are ${\mathcal L}_0=88\pm 1$ ppm, $A=346 \pm 4$ ppm, and $\tau=1490 \pm 75$ days. As the modified $\chi^2/{\rm ndf}=1.6$ for both linear and exponential fits, only the result of linear fit is discussed in the text. Vertical grid lines correspond to the first day of the year shown in the tick label. Typical atom position is $Z=1.4$ mm until July, 2017, and after that, various distances are used ranging from $Z=0.14$ mm to 1.4 mm, settling down to $Z=0.42$ mm in February 2018. } 
 \label{FvsT}
\end{center}
\end{figure}

Although the cloud of atoms is located relatively close to the mirror surface, the deterioration of the quality of the mirror is slow, as Fig. \ref{FvsT} shows. Particularly important is the fact that the rate of the mirror loss increase did not change even after we moved the atoms closer to the mirror surface in the middle of 2017. This suggests that the mechanism of the increase in the loss is not atoms coating the mirror surface, and hence, the possibility of fitting the data with a single curve. The rate 6.76(1) ppb/h of the mirror loss increase obtained by the fit is similar to the room-temperature case in Ref. \cite{OptExp.23.18014}, which reports measurements on a test setup with two mirrors with a large ROC and without atoms to characterize the degradation of high-reflectivity mirrors in vacuum (see Table \ref{LossComp} for comparison). As for the same top layer material of SiO$_2$, 100 $^{\circ}$C data in Ref. \cite{OptExp.23.18014} is converted to 1 ppb/h at room temperature, following Fig. 3 of Ref. \cite{OptExp.23.18014}, which is of the same order of magnitude as the measured rate of the loss increase. Other systems we can compare with are those in Ref. \cite{PhysRevA.95.033812} and Ref. \cite{1902.06014}, both of which are reports on cQED systems for Yb$^+$ ions. Our system has a significantly lower rate of the loss increase compared to that in Ref. \cite{PhysRevA.95.033812}, but higher than that in Ref. \cite{1902.06014}. This is consistent with the hypothesis discussed in Ref. \cite{1902.06014} that UV-enhanced deposition of hydrocarbons onto the cavity mirrors degrades the finesse. 
In spite of this larger increase in the loss than Ref. \cite{1902.06014}, the rate of the loss increase is still low enough to maintain the high finesse over the time scale of many years. This atomic ensemble near a mirror surface without excessive contamination is suitable for performing a wide variety of cQED experiments in the strong coupling regime, where $\eta>1$.

\section{Conclusion}
We have demonstrated the loading of $^{171}$Yb atoms into a tight one-dimensional optical lattice in an optical cavity close to the surface of a mirror. Two distinct regimes of efficient loading of atoms from a mirror MOT into the optical lattice are observed, due to two different loading mechanisms. The loading of atoms in the lattice is performed simply by putting a MOT at a desired location, and up to 1500 atoms are trapped in an optical lattice with a 6.6-$\mu$m waist at a distance of 0.14 mm from the mirror surface. These results open a simple way to realize a system suitable for quantum mechanical manipulation of atoms in the strong-coupling regime. 

\begin{acknowledgments}
This work was supported by DARPA, NSF, NSF CUA, and ONR. B.B. acknowledges the support of the Banting Postdoctoral Fellowship. A.K. acknowledges the partial support of a William M. and Jane D. Fairbank Postdoctoral Fellowship of Stanford University.
\end{acknowledgments}

\bibliographystyle{apsrev4-1}
\bibliography{Trapping}

\end{document}